# Ab initio study on optimization of bisphenol A and phosgene to manufacture polycarbonate

Bahtiyar A. Mamedov, Elif Somuncu, Ebru Karatas,

Department of Physics, Faculty of Arts and Sciences, Gaziosmanpaşa University, Tokat, Turkey

E-mail: bahtiyar.mehmetoglu@gop.edu.tr; elf_smnc@hotmail.com; karatas08@gmail.com

**Abstract**

Geometric optimization is played an important to manufacture and design materials in many implementations. Therefore the choice of optimization method is of considerable significance to easily solve the problems. In this work, the ground state geometries for bisphenol A and phosgene that manufacture polycarbonate have been optimized using the Hartree-Fock method with different basis sets. The optimization results for bisphenol A and phosgene are compared with theoretical and experimental data. The obtained optimizations results have been show that our data are in agreement with the literature and experimental data.

## 1. Introduction

Polycarbonate (PC) is a thermoplastic material in which presents producer opportunities for design simplicity and cost decrease [1]. It is a hard, transparent, and amorphous thermoplastic polymer dependent together by carbonate groups and presents an unmatched compound of properties [1-3]. PC is requested commonly in the plastic industry which is a highly clear plastic that may transmit over 90% of light as well as glass, resistant to UV light, and transparent [4]. Therefore PC material is used to make eyewear [4-5]. Many researchers have been manufactured the PC by using condensation polymerization of bisphenol A and phosgene [6-9]. Nowadays, it is still worked for evaluation precisely and correctly the physical and chemical properties of bisphenol A and phosgene [10]. Therefore, many theoretical method such as ab initio methods for the determination of these properties of bisphenol A and phosgene have been suggested in the literature. One of ab initio methods is Hartree Fock (HF) method. This method is an approximation proposed for the evaluation of the wave function and the energy of a quantum many-body system in computational physics and chemistry [11].

In this work, the ground state geometries of bisphenol A and phosgene have been optimized by using the HF method with 3-21G, 3-21+G, 6-31G, 6-31+G, 6-31+(d), 6-311G, and 6-311+G basis sets. The optimization results of bisphenol A and phosgene are compared with literature data. The optimizations results are in agreement with the theoretical and experimental data.

## 2. Method

The ground state geometries of bisphenol A and phosgene are optimized using the HF with different basis sets without that symmetry limitation in the molecules. Since our aim focuses on the accurate evaluation of ground state geometries optimized, we have used the HF method applying over 3-21G, 3-21+G, 6-31G, 6-31+G, 6-31+(d), 6-311G, and 6-311+G basis set. All calculations have been made using the Gaussian 05 program package. The interatomic bond lengths and bond angle have been compared with experimental data.

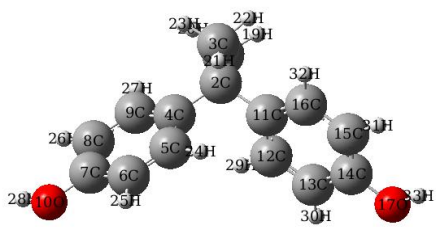

a)

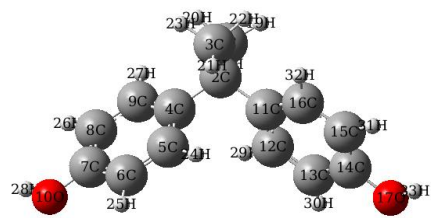

b)

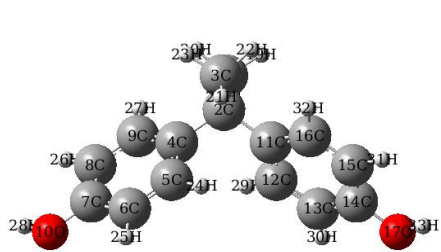

e)

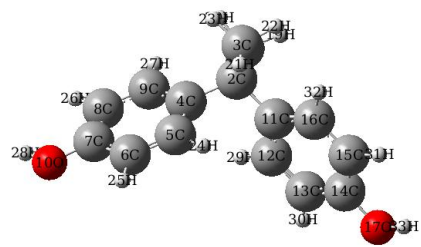

d)

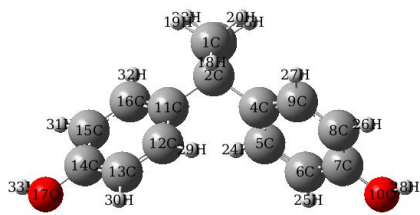

e)

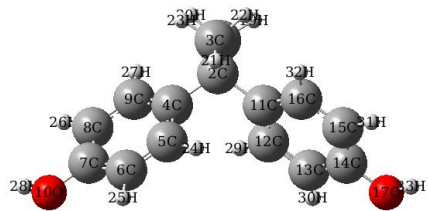

f)

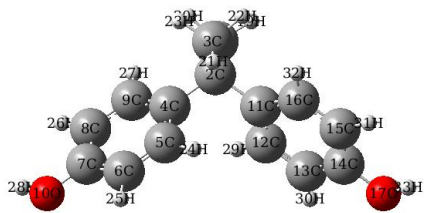

g)

**Figure 1.** Optimized geometries of bisphenol A a) HF/3-21 b) HF/3-21+G  c) HF/6-31   d) HF/6-31+G  e) HF/6-31+G (d)  f) HF/6-311 g) HF/6-311+G

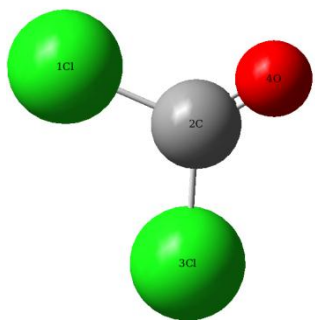

a)

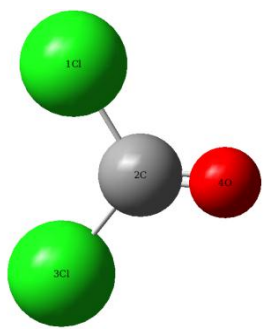

b)

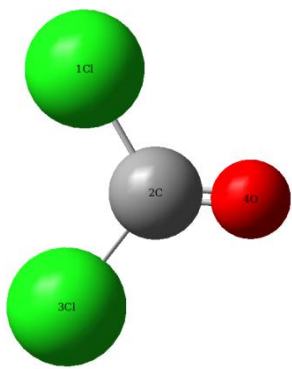 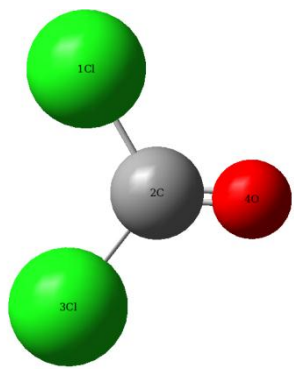

c)                                                 d)

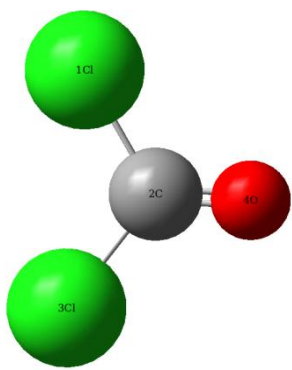 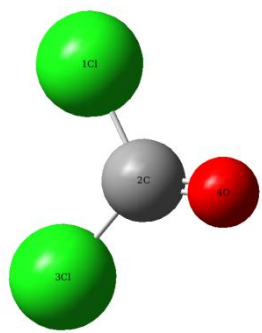

e)                                                 f)

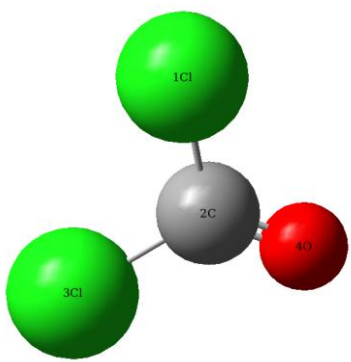

g)

**Figure 2.** Optimized geometries of phosgene a) HF/3-21 b) HF/3-21+G c) HF/6-31   d) HF/6-31+G  e) HF/6-31+G (d)  f) HF/6-311 g) HF/6-311+G

Table 1 Optimization values of model molecules poycarbonate for different ab initio methods and basis set

| Parameter | HF/3-21 | HF/3-21+G | HF/6-31 | HF/6-31+G | HF/6-31+G(d) | HF/6-311 | HF/6-311+G | Experiment [12,13] |
|---|---|---|---|---|---|---|---|---|
| Bond length ($\overset{o}{A}$) | | | | bisphenol A | | | | |
| C(2)-C(11) | 1.537 | 1.5383 | 1.5414 | 1.5417 | 1.5409 | 1.541 | 1.5412 | 1.536 |
| C(2)-C(4) | 1.5371 | 1.5383 | 1.5414 | 1.5417 | 1.5409 | 1.541 | 1.5412 | 1.536 |
| Bond angle ($^o$) | | | | | | | | |
| C(4)-C(2)-C(11) | 109.6627 | 109.2251 | 109.9082 | 109.9123 | 109.9848 | 109.919 | 109.915 | 109.2,109.5 |
| Bond length ($\overset{o}{A}$) | | | | phosgene | | | | |
| C(2)-O(4) | 1.1644 | 1.1729 | 1.1751 | 1.1743 | 1.1601 | 1.1714 | 1.1709 | 1.1785 |
| C(2)-Cl(1) | 1.8303 | 1.8155 | 1.7932 | 1.7909 | 1.7335 | 1.7891 | 1.7868 | 1.7424 |
| Bond angle ($^o$) | | | | | | | | |
| O(4)-C(2)-Cl(1) | 124.9135 | 123.818 | 123.4842 | 123.5163 | 123.3833 | 123.5758 | 123.5944 | 124.09 |
| Cl(1)-C(2)-Cl(3) | 110.1729 | 112.364 | 113.0316 | 112.9673 | 113.2334 | 112.8485 | 112.8111 | 111.83 |

## 3. Numerical Results and Discussion

The basic goal of this paper is to examine the optimization using the HF method with different basis sets and compare the results available with literature. Optimization of the bisphenol A and phosgene by using the HF over 3-21G, 3-21+G, 6-31G, 6-31+G, 6-31+(d), 6-311G, and 6-311+G basis set is presented in this study. The optimization of the bisphenol A and phosgene was calculated using Gaussian 0.5 packet program. In order to demonstrate that the results of the calculation are accurate and precise, we have compared with experimental data. The obtained optimization results for bisphenol A and phosgene is given in Table 1. As seen in Table 1, the obtained optimization results give values very closer to the experimental data [12-13]. The results obtained optimization by using the HF over 3-21G for bisphenol A are in better agreement with the experimental data than other studies [12]. Thus our results show that the optimization of bisphenol A and phosgene are correct and sensitive. Also optimization geometries are given in Figures 1-2. The preciseness and accurateness of the obtained optimization results are suitable and can be suggested for the reaction to be composed of bisphenol A and phosgene. The novelty of this study is that the present work is more efficient and performed accurately for the optimization of bisphenol A and phosgene in the HF method over 3-21G, 3-21+G, 6-31G, 6-31+G, 6-31+(d), 6-311G, and 6-311+G basis set.

## Conclusion

As known, polycarbonate, which occurs from bisphenol A and phosgene, is a significant polymer with a variety of optical applications such as glasses, lenses, and laboratory safety goggles. Therefore, optimization of bisphenol A and phosgene will be useful to evaluate the physical and chemical properties of these products manufactured polycarbonate material.